\documentclass[a4paper]{article}
\usepackage{INTERSPEECH2019}
\usepackage[T1]{fontenc}
\usepackage[utf8]{inputenc}
\usepackage{tipa}
\usepackage{tipx}
\usepackage{amsmath}
\usepackage{amssymb}
\usepackage[table]{xcolor}
\usepackage{url}
\usepackage{colortbl}

\DeclareUnicodeCharacter{2D0}{\textipa{:}}
\DeclareUnicodeCharacter{292}{\textipa{Z}}
\DeclareUnicodeCharacter{283}{\textipa{S}}
\DeclareUnicodeCharacter{279}{\textturnr}
\DeclareUnicodeCharacter{26A}{\textipa{I}}
\DeclareUnicodeCharacter{251}{\textipa{A}}
\DeclareUnicodeCharacter{254}{\textipa{O}}
\DeclareUnicodeCharacter{1EBD}{\textipa{\~e}}
\DeclareUnicodeCharacter{28C}{\textturnv}
\DeclareUnicodeCharacter{259}{\textreve}
\DeclareUnicodeCharacter{28A}{\textupsilon}  
\DeclareUnicodeCharacter{25B}{\textepsilon} 
\DeclareUnicodeCharacter{28F}{\textscy}     
\DeclareUnicodeCharacter{289}{\textbaru}    
\DeclareUnicodeCharacter{275}{\textbaro}    
\DeclareUnicodeCharacter{252}{\textturnscripta} 
\DeclareUnicodeCharacter{25D}{\textipa{\textepsilon\textrhoticity}} 

\DeclareUnicodeCharacter{27E}{\textfishhookr} 
\DeclareUnicodeCharacter{2DE}{\textrhoticity} 
\DeclareUnicodeCharacter{25A}{\textschwa\textrhoticity} 
\DeclareUnicodeCharacter{255}{\textctc} 
\DeclareUnicodeCharacter{2B0}{\textsuperscript{h}}

\DeclareUnicodeCharacter{329}{\textsyllabic{ }} 

\DeclareUnicodeCharacter{3B8}{\texttheta} 
\DeclareUnicodeCharacter{26D}{\textrtaill} 
\DeclareUnicodeCharacter{25C}{\textrevepsilon} 
\DeclareUnicodeCharacter{282}{\textrtails}  

\DeclareUnicodeCharacter{267}{\textrtailhth} 

\DeclareUnicodeCharacter{256}{\textrtaild}
\DeclareUnicodeCharacter{273}{\textrtailn}
\DeclareUnicodeCharacter{14B}{\textipa{N}} 

\DeclareUnicodeCharacter{288}{\textrtailt}
  
\DeclareUnicodeCharacter{26B}{\textipa{\textltilde}} 

\title{
Transparent pronunciation scoring using articulatorily weighted phoneme edit distance}
\name{Reima Karhila$^1$, Anna-Riikka Smolander$^2$, Sari Ylinen$^2$, Mikko Kurimo$^1$}
\address{
  $^1$Aalto University School of Electrical Engineering\\
  $^2$University of Helsinki}
\email{firstname.lastname@aalto.fi, firstname.lastname@helsinki.fi}

\begin{document}

\maketitle
\begin{abstract}
For researching effects of gamification in foreign language learning for children in the “Say It Again, Kid!” project we developed a feedback paradigm that can drive gameplay in pronunciation learning games. We describe our scoring system based on the difference between a reference phone sequence and the output of a multilingual CTC phoneme recogniser.  We present a white-box scoring model of mapped weighted Levenshtein edit distance between reference and error with error weights for articulatory differences computed from a training set of scored utterances. The system can produce a human-readable list of each detected mispronunciation's contribution to the utterance score. We compare our scoring method to established black box methods.

\end{abstract}

\section{Introduction}

In Computer Assisted Pronunciation Training (CAPT) our goal is to give speakers feedback on how to improve their pronunciation skills. 
With the assumption that mispronunciations are detectable as phoneme detection or classification errors,
we can apply variants derived from Automatic Speech Recognition (ASR) acoustic models to evaluate pronunciation skills of a speaker. 
For example mispronunciation detection with Extended Recognition Networks (ERN)~\cite{harrison2009implementation} or Goodness of Pronunciation (GOP) scoring based on posterior likelihoods~\cite{witt2000phone,kanters2009goodness} 
are based normal ASR acoustic models.
%
%
Deep Neural Network (DNN) based Acoustic modelling has improved ASR performance vastly in the recent decade 
 and DNN models form an excellent basis for new experiments in CAPT tasks.

%
%

Our CAPT game “Say It Again, Kid!"~\cite{Karhila2017,ylinen2017} is used to study children's foreign language acquisition. We 
have developed a feedback mechanism with constraints set by the project:

(1)    \emph{The feedback must be usable as a gaming element:} We need a 0-5 star feedback score for each individual utterances. We also need to compute the score quickly, so we use single pass unidirectional processing.
    
(2) \emph{The system must produce meaningful analytics for teachers and researchers:}
    We need statistics on what phones are difficult for a speaker and what has been learned during game play. As International Phonetic Alphabet (IPA) is widely known among teachers, we report mispronunciations in IPA characters.

(3) \emph{We do not have resources to generate phonetic transcriptions of pronunciation irregularities in L2 data:} Our analytics system is trained on large native databases and the scoring system is trained on  a collection of roughly scored utterances.
    
(4) \emph{The single utterance annotation scores are not reliable:} The scoring system must be robust so it can be trained on noisy annotations, and must produce scores that the users find justified.

Describing utterances as chains of articulatory or phonological feature vectors has been a popular approach for mispronunciation detection~\cite{7820800, duan2017effective, arora2017phonological, arora2018phonological}. 
Noting that any realisable articulatory feature vector can be represented by a phonetic unit from the IPA alphabet, we settled on a simpler phone-based approach, extracting articulatory features by mapping phones to feature vectors.
For L2 CAPT, accurate scoring based on detected phonemes requires a recogniser that covers the phones of the target language as well as the phonetic space for mispronunciations. To extend the phone set, the L1 of the speakers is a natural choice, but including more languages will cover a larger space of mispronunciations and decrease the L1 dependency.

The Connectionist Temporal Classification (CTC)~\cite{graves2006connectionist} is an end-to-end ASR framework that models all aspects of the speech sequence -- both acoustic and linguistic aspects. It does not require segmentation of its training inputs nor does it produce phone alignments in the same way as Hidden Markov Model (HMM) based models do.
Duration information is essential for language learning for languages where phone duration is important in distinguishing words from others. As our CAPT game 
aims to teach a  standardised way of pronunciation,
feedback for phoneme duration is valuable also in languages, where duration does not have such a phonemic role.
As phone duration is encoded within the network itself 
and can be learned from native speech samples directly, CTC provides a good base for producing duration based feedback.

We can use the list of string edit operations that show the difference between the phonetic forms of scored and reference utterance as analytical feedback. Additionally we can use the list to compute  Phonologically Weighted Levenshtein Distance (PWLD)~\cite{fontan2016using} based error, where the effect of every mispronunciation is fully transparent, and use it as a basis for a point-based feedback score.
By computing error costs for phones based on articulatory/phonological feature weights shared between all phones, we can train a robust low-dimensional regression system from a small number of annotated L2 learner samples.

In this paper we present an experiment on scoring utterances in a L2 CAPT game, using multilingual CTC phoneme recognition and a data-driven variant of PWLD.
There are few academic reports on 
CAPT in learning games, 
we found
only one~\cite{young2014game}.
And generally work on scoring of L2 utterances has been done on in-house data sets and reported in statistical manner. 
Though we are unable to share our audio data, by sharing our recognition results and scoring experiments as an online compendium\footnote{The online compendium can be found at \url{https://github.com/aalto-speech/interspeech2019_karhila_et_al}}
 to the paper, we hope to give some concrete samples for the field of automatic scoring of L2 learners' utterances.
%

\section{Scoring based on PWLD}
\label{sec:scoring}

The Levenshtein distance used to compute error rates for ASR evaluation treats all insertion, deletion and substitution operations identically with a unit cost of 1.
Phoneme error rate based on PWLD has been used for automatic prediction of intelligibility in~\cite{fontan2016using}, where the substitution weights are based on the number of differing articulatory features between each phone pair.

We propose using a corpus of L2 utterances with annotated scores to optimise weights for PWLD to form Data Driven Phonologically Weighted Levenshtein Distance (DD-PWLD). 
In our approach also insertion and deletion costs are computed based on articulatory features.
Thus the edit distance is computed as 
\begin{equation}
\boldsymbol{D} = \boldsymbol{O}_{sub}\boldsymbol{F}_{sub} + \boldsymbol{O}_{ins}\boldsymbol{F}_{ins} + \boldsymbol{O}_{del}\boldsymbol{F}_{del} 
\end{equation}
where 
$\boldsymbol{O}_{sub}$, $\boldsymbol{O}_{ins}$ and $\boldsymbol{O}_{del}$ are $ n \times p $ vectors of substitution, insertion and deletion operations for each utterance,
%
$\boldsymbol{F}_{sub}$, ${F}_{ins}$ and $\boldsymbol{F}_{del}$ are the $f$-dimensional vectors of articulatory error costs for substitution, insertion and deletion costs.
The operation matrices $\boldsymbol{O}$ are composed from edit operation lists 
of the hypotheses with the smallest errors in $N$-best hypothesis lists for the utterances. 
The mapping to a score in the range $[0,r]$ is
\begin{equation}
 \boldsymbol{ \hat y} = r \Big(1 - \tanh \Big( a  
  \frac{\boldsymbol{D}}
       {\boldsymbol{L}^l}\Big)\Big)
\end{equation}
where 
$\boldsymbol{L}$ is a vector representing phoneme counts of the reference utterances, $l$ is the length compensation variable, $a$ is general scaling for the mapping,
$\boldsymbol{y}$ the human annotations for the utterances. 
We want to choose  $\boldsymbol{F}_{sub}$, 
$\boldsymbol{F}_{ins}$, 
$\boldsymbol{F}_{del}$, 
$l$ and 
$a$ to minimise the error between predictions 
$\boldsymbol {\hat y}$ and reference annotations.

\begin{figure}
  \centering
  \includegraphics[width=0.45\textwidth]{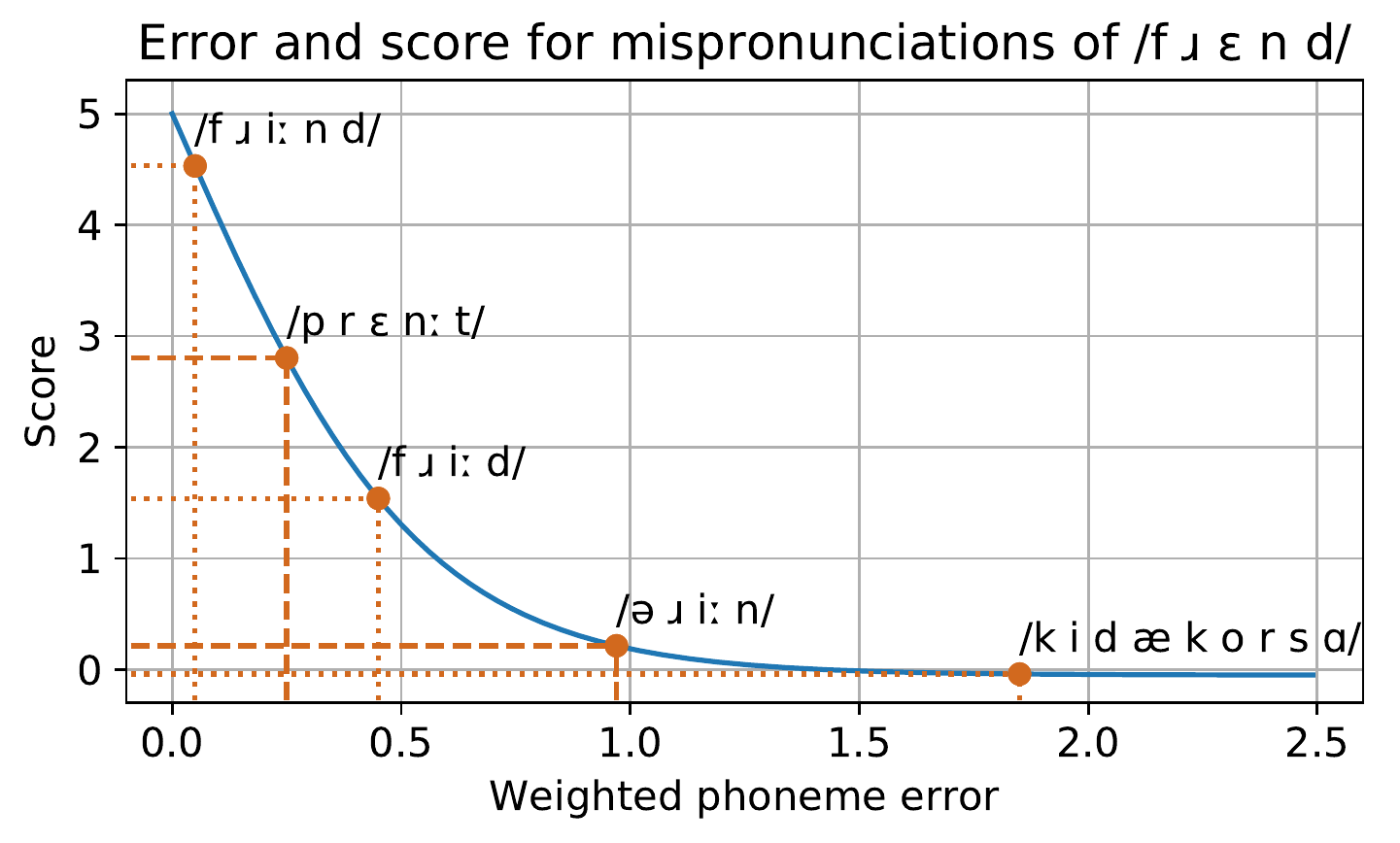}
  \caption{Phoneme recognition sample results, their edit distances and scores.}
  \label{fig:error_and_mapped_score}
\end{figure}

\begin{table}
  \caption{Sample breakdown of scoring for a pronunciation attempt using the transparent scoring scheme in a format that can be shown to a user or a teacher.}
  \begin{center}
    \framebox[0.4\textwidth]{
      \begin{tabular}{llll}
        \multicolumn{2}{l}{ Target pronunciation: } & \multicolumn{2}{l}{/\emph{f ɹ ɛ n d}/} \\
        \multicolumn{2}{l}{ Your attempt: } & \multicolumn{2}{l}{/\emph{p r ɛ nː t}/ } \\
        \multicolumn{2}{l}{ Total errors: } & \multicolumn{2}{l}{ 2.3 } \\
        \multicolumn{2}{l}{ Length compensation multiplier: } & \multicolumn{2}{l}{  6.4 } \\
        \multicolumn{2}{l}{ Mapped score: } & \multicolumn{2}{l}{ 2.8 } \\
        \multicolumn{4}{l}{ Breakdown: } \\
        /\emph{f}/ $\rightarrow$ /\emph{p}/ & \multicolumn{2}{l}{Plosive instead of fricative} & -0.85 \\
        /\emph{ɹ}/ $\rightarrow$ /\emph{r}/ & \multicolumn{2}{l}{Trill instead of approximant} & -0.63 \\
        /\emph{n}/ $\rightarrow$ /\emph{nː}/ & \multicolumn{2}{l}{Too long}  & -0.44 \\
        /\emph{d}/ $\rightarrow$ /\emph{t}/ & \multicolumn{2}{l}{Not voiced} &  -0.38 \\
      \end{tabular}
    }
  \end{center}
  \label{tab:sample_breakdown}
  \vspace{-0.5cm}
\end{table}

Samples of the mapped scores for different mispronunciations of the utterance "friend" are shown in Figure~\ref{fig:error_and_mapped_score}.
The scoring can be broken down as described by the list of edit operations that is a byproduct of the weighted Levenshtein
distance.
One sample analytical scoring is shown in Table~\ref{tab:sample_breakdown}.

\section{Experiments}
\label{sec:experiment}

We present an experiment of scoring L2 learner utterances where
target language is UK English and the speakers have a Finnish background.
Recognition results and scoring training utilities are available
at the compendium repository.

\subsection{Data}

UK English as the target language is the most important.
This is covered by WSJCAM0 corpus for adult speech and PF-STAR for children's speech.
Finnish as a the native
language of the language learner data set is also important,
and is covered by the Finnish SPEECON and SPEECHDAT databases,
both of which contain both adult and child speech.
%


As the CAPT game aims for UK pronunciation, we added  American English to the training pool to better detect non-UK pronunciations.
This was easily available in WSJ1 for adults and TI-DIGITS for children (adult data from TI-DIGITS was not used).
As our aim is
to cover as much of the IPA acoustic-phonetic
space for non-tonal languages as possible using the resources easily available to us,
%
we added Swedish from Spraakbanken as Swedish includes vowels and retroflex consonants that do not exist in the other languages available to us.
This is by far not a complete coverage of the IPA space, but is a solid start
for our current purposes.






L2 UK English data 
was collected from Finnish 10-12 year old children using the game prototype. Additional game data was collected from native English speakers of the same age group with Southern English dialect.

The dictionaries used are CMUdict for American English,
COMBILEX for American and UK English~\cite{fitt2006redundancy},
Språkbanken for Swedish 
 and a combination of Speechdat lexicon and rule-based mapping
for Finnish.

\subsection{Annotations}

Annotations were made by a single annotator.
As it is generally known that the reliability of labels produced by a single annotator is questionable, we pay attention extra attention to downplay outlier influence when training our scoring system in Section~\ref{sec:scoring}.



The annotations were first made on a 0-100 scale and later mapped to the
0-5 scale.
The utterances were grouped by reference phrase. 
The annotator listened to all the samples in a group, and tried to find samples that represent the best pronunciations and samples that represent typical errors. 
%
For most utterance groups, there were one or more mispronunciation patterns that showed up repeatedly 
These were usually one phoneme away from the 
model pronunciation. If the attempt differed from the model with 
one feature e.g. voicedness, the score was deducted from 100 to 90. If the 
attempt differed more than that the reduction 
was bigger, depending on the distance of articulation place or 
manner.  This kind of reduction according the phonemes was 
possible with the 
best pronounced samples. The more mistakes 
there were, the more difficult it was to calculate the labelling 
with such a simple system. 


%
This scoring framework was possible as the samples were mainly one-word long. With longer sentences the labelling was more complicated and in these  the main focus was on the words that were pronounced mostly correctly. 
Because the samples were short, prosody was not used as a criterion
in the labelling. 
Even though e.g. stress was in many cases foreign, it mostly did not reduce the intelligibility and clarity of the phones.


\subsection{Phoneme set}
\label{lab:phoneme_set}

The phone set consists of all the phones used in the various
pronunciation dictionaries of the four languages or dialects.
The grouping of phones between languages is based on the
IPA alphabet. 
Some phones are combined to form more meaningful units for pronunciation learning. For example, all diphthongs and many vowel sequences like /\emph{iø}/ and /\emph{iɑ}/ as well as some combined consonants like /\emph{\texttoptiebar{dʒ}}/ are represented as single units.
The phonetic units are listed in 
the online compendium. Of the 157 different phonetic units used, 28 are unique to one language, the rest are shared between at least two languages or dialects.
The dictionary mappings and phone combinations are evolving work, and the list should not be used as a definite reference. For example, the geminated /\emph{\texttoptiebar{tːs}}/ present in the Finnish language is represented by a single unit, but the more ubiquitous normal length variation /\emph{\texttoptiebar{ts}}/ is represented by /\emph{t}/ and /\emph{s}/ separately.
Also rhoticity at the end of vowel sequences needs a closer look.
Future versions will see these improved.

\subsection{Models}

Three model sets were used in the experiment.
A HMM-GMM model is trained to select correct pronunciation variants and produce alignments, a CTC model is trained to do phoneme recognition, and a regression model to turn the recognition outputs into scores.
%
%

\subsection{HMM-GMM training}

\begin{table}[t]
  \begin{center}
    \caption{Error rates for the recognition systems. These \textbf{should not be compared to the state-of-the-art,} as the only function for the HMM-GMM models is alignment. What the results show is that the extended phoneme set is not ideal for basic recognition tasks and how the performance drops when speech data is pooled across languages. WER for HMM-GMMs is computed on test sets, the PER for CTC on development sets.}
  \label{tab:performance_drop}
  \begin{tabular}{|lccc|c|}
  \hline
    \bfseries  & \bfseries HMM  & \bfseries HMM & \bfseries  & \bfseries CTC  \\
    \bfseries  & \bfseries Unmrg. & \bfseries  Merged & \bfseries Rel. & \bfseries Merged \\
    \bfseries Data set & \bfseries WER & \bfseries  WER & \bfseries  change & \bfseries PER* \\
    \hline
    \multicolumn{5}{|l|}{UK ENGLISH} \\
    PF-star & 7.83 & 14.72 & +87\% & 55.2 \\
    WSJ0 si mic1 & 26.47 & 31.43 & +18\% & 51.5 \\
    WSJ0 si mic2 & 27.68 & 32.28 & +17\% & 51.4 \\
    \hline
    \multicolumn{5}{|l|}{US ENGLISH} \\
    WSJ1 si mic1 & - & 31.43 & - & 51.5 \\
    WSJ1 si mic2 & - & 26.21 & - & 51.0 \\
    TIDIGITS  & - & 0.83 & - & - \\
    \hline    
    \multicolumn{5}{|l|}{FINNISH} \\
    Speechdat & 7.60 & 8.54 & +12\%& 45.8 \\
    Speecon & & & & \\
    - clean mic0 & 6.07 & 7.72 & +27\% & 48.0 \\
    - clean mic1 & 8.41 & 10.47 & +24\% & 51.0 \\
    - café mic0 & 6.98 & 7.72 & +11\% & 50.1 \\
    - café mic1 & 7.55 & 9.70 & +28\% & 49.2 \\
    - children mic0 & 11.69 & 12.90 & +10\% & 48.1\\
    - children mic1 & 12.83  & 15.45 & +20\% & 48.2 \\
    \hline
    \multicolumn{5}{|l|}{SWEDISH} \\
    Spraakbanken  & - & 41.46 & - &  53.4 \\    
    \hline
  \end{tabular}
  \end{center}
  \vspace{-0.5cm}
\end{table}

The speaker-adaptive HMM-GMM model is trained with Kaldi~\cite{Povey_ASRU2011}.
It is only used to to choose best pronunciation alternatives and provide segmentations
for the training utterances that are fed into the CTC training.
Several model sets are trained. First, a combined model set trained with all the data is used
for segmenting the training data. This is trained with long and short variations of phonetic units
merged.
Another model set is trained only on languages where duration information is phonetically relevant.
Where it matters, each segment of training utterances as segmented by the combined system is postprocessed
to be short or long variation.
This is more normative than descriptive. Our interest lies in guiding the users of our system to
attune the durations of their uttered phones towards some standard, which becomes thus defined
by the data. By including the duration information in the model set, we enable the CTC recognition
network to learn it.

Table~\ref{tab:performance_drop} shows the reduction in performance caused by the merged phone set.
For most of the test sets, language models (LM) used for the decoding are trained from training
data labels. 
The table shows that data pooling over languages degrades performance. It also shows that performance for close microphones is better than table microphones.

\subsection{CTC model training}

The CTC model is a 3 layer deep and 600 units wide Gated Recurrent Unit (GRU) Recurrent Neural Network (RNN) running on Tensorflow~\cite{tensorflow2015-whitepaper}.
The model is trained with batches of 64 randomly cut 2 second segments with random resampling between 0.95 and 1.05 speed. A random linear sum of noise signals is added to the samples~\cite{hannun2014deep} and the final waveform is filtered by a random filter to emulate microphone differences~\cite{Raesaenen2018comparison}.
The reference labels are subsections of label strings and are cut based on time alignments produced by basic HMM-GMM model.
 40 Mel-weighted spectral bins computed on 512 point FFT on 25 ms frames at 10 ms interval on 8 kHz audio were computed and appended with a binary vector of 
Speaker age groups (-5, 6-8, 9-12, 13-17, 18-64, 65-) and gender.
Table~\ref{tab:performance_drop} shows the CTC Phoneme Error Rate (PER) performance. 
Already it performs surprisingly consistently for all the native data sets, getting around 50\% of the phones right, and the performance drop between close and table microphones is much smaller than with the HMM-GMM systems.

\subsection{Articulatory Features}

Each phone is described by an articulatory feature vector.
Our articulatory feature set is based on question sets used in model clustering parametric speech synthesis.
Additional features were added for diphthongs and vowel sequenecs, describing the direction of movement of the articulation place.
Table~\ref{tab:art_foo} lists the features used in this work.
Compared to the 18 articulatory parameters in~\cite{arora2018phonological} or 14  in~\cite{fontan2016using}, our selection of 55 descriptors is substantially larger and can cover a much larger segment of the IPA space without duplicate mappings between phones and feature vectors. 

\begin{table}[t]
  \begin{center}
  \caption{Articulatory features used in this work}
  \label{tab:art_foo}
  \begin{tabular}{|p{8cm}|}
    \hline
    
 \multicolumn{1}{|c|}{\bfseries Articulatory features for vowels} \\
 dipthong, long, rhotic, unround-schwa \\
 front, nearfront, central, nearback, back, \\
 open,  nearopen, openmid,  mid, closemid, nearclose, close, \\
 rounded, unrounded, \\
 diphthong-forward, diphthong-backward, \\
 diphthong-opening, diphthong-closing, \\
 diphthong-rounding, diphthong-unrounding,\\
    \hline
 \multicolumn{1}{|c|}{\bfseries Articulatory features for non-vowels} \\
 affricate, approximant, fricative, plosive, nasal, trill, \\
 alveolar, bilabial, coronal, dental, dorsal, labial, labiodental,\\
 lateral, postalveolar, velar \\
 pulmonic, retroflexed, syllabic, palatalized, aspirated, \\
 lenis, fortis,
 labialized, voiced, unvoiced, geminated \\
\hline
  \end{tabular}
  \end{center}
      \vspace{-0.5cm}
\end{table}

\subsection{Scoring}
\label{sec:scoring}
Scoring was done by two linear regressors (LR) and two non-linear ones. All optimisations are done on the L2 speech database, with training data used once with real labels and scores, and once with shuffled labels and rejected scores.
One tenth of the training data is split to be the development set.
PWLD-LR uses the default phonological weights and only terms $a$ and $l$ used for mapping the distance to a score are optimised. DDPWLD-LR additionally optimises the phonological weights $\boldsymbol{F}$ for computing the distance.
Optimal values are found with the BFGS optimiser. 
To reduce outlier influence in optimisation, Cauchy norm (using log of error) and 
appropriate bounds for values are used.
%
The initial best paths from the $N$-best list are computed with the cost presented in~\cite{fontan2016using}. 
Support Vector Machine Regressors (SVR) and Random Forest Regressors (RF) are trained on the same parameterisations as the PWLD, concatenated 
with the length of the reference utterance in number of phones.
Robust parameters are selected using the separate development set.

\subsection{Results}

Table~\ref{tab:score_corr} lists the correlations between human annotations and predictions. The left result column shows the correlation for all data and the right column shows the results when outliers have been removed from the test set. If at least two of the regressors produced an error more than 2 standard deviations above their mean error, the data item in question was marked as an outlier. Out of the 4313 test items, 205 were considered outliers.
Compared to PWLD, data-driven estimation of weights for DDPWLD improved the performance by around 50\%, almost to the level of SVM and RF performance. The black box methods still beat the DDPWLD by a margin, but the drop in performance is compensated 
 in DDPWLD by the ability to assign a cost for every individual operation.


\begin{table}[t]
  \begin{center}
    \caption{Correlation between predictions and human annotations for the scoring test set.}
  \label{tab:score_corr}
  \begin{tabular}{|lcc|}
  \hline
      \bfseries&  \bfseries  & \bfseries Outliers \\
    \bfseries System & \bfseries All & \bfseries removed \\
    \hline
    PWLD-LR & 0.31 & 0.36 \\
    DDPWLD-LR & 0.47 &  0.54 \\
    PWLD-SVM & 0.52 & 0.61 \\
    PWLD-RF & 0.49 & 0.56 \\
    \hline
    \end{tabular}
    \end{center}
    \vspace{-0.5cm}
\end{table}

\section{Discussion}
\label{sec:discussion}

The description of errors as string operations and describing the sum of string operations as vectors of articulatory differences between reference and hypothesis is an efficient way of parameterising pronunciation mistakes.
It gives a robust basis for a scoring system. The system's leniency is dependent on the quality of the underlying phoneme recogniser. Most recognition errors reduce the score rather than improve it. Therefore the worse the recogniser, the harder it is to get a high score. At the moment the recogniser is performing at around 50\% phoneme error rate for the native test sets, which is still considerably higher than the 30\% letter error rates reported for the original CTC setups~\cite{graves2006connectionist}. By using $N$-best lists, the system works well enough for deployment in a game. As mentioned in Section~\ref{lab:phoneme_set}, the phoneme merging between the data sets is
not finished and some performance gains are expected from correcting errors there.

Another problem lies mostly on the recogniser's side.
Allowing mid-utterance code-switching is difficult with only native utterances in the training data.
The CTC network relies too much on past information and is not able to correctly find phoneme sequences that contain phone sequences unique to different languages.
An example is a mispronunciation of the Finnish word ``korkea'' /\emph{korkea}/
with the trill /\emph{r}/ replaced by an alveolar approximant /\emph{ɹ}/ results in /\emph{koɹkʰa}/. Here, the vowel sequence /\emph{ea}/, which is foreign to UK English, is replaced by /\emph{kʰa}/. 
Another is a mispronunciation of the English word ``friend''  /\emph{fɹɛnd}/ with the alveolar approximant replaced by a trill resulting in the recognition as /\emph{prɛnːt}/.
As the phone combination /\emph{fr}/ is not adequately presented in the training data, 
the whole utterance is recognised as if it were in the Finnish phonetic space.

The quality of the test data gathered from language learners using an early version of the CAPT data is very variable in speaking and recording quality, and 
the recordings can start and stop in the middle of the utterance.
%
%
The code switching issue needs to be investigated. The fact that the GRU internal states are reset to zero for every training batch probably makes the system less prone to code-switching. Reusing the internal state as is for the next batch might also be problematic, as the extra speaker information is included at a low level. A review of best practises needs to be done.

Word boundaries are also problematic. For example the Finnish utterance, ``on koala'' is pronounced either /\emph{on~koɑlɑ}/ or /\emph{oŋkoɑlɑ}/, but always with a vowel sequence /\emph{oɑ}/, whereas the utterance ``onko ala?'' is pronounced either as /\emph{oŋko~ɑlɑ}/ or /\emph{oŋkoɑlɑ}/, with the /\emph{oɑ}/ combination either as vowel sequence or as slightly separated phones.


\section{Conclusions}
\label{sec:conclusion}

We have presented a rather simple two-component system for scoring L2 learner utterances using a CTC phone recogniser and DDPWLD as an error measure.
We found that this simple linear scoring method performs almost as well as the established black box regression methods.

Future work includes improving the integration of the different phoneme sets and  dictionaries and develop further our phoneme recogniser.
An in-depth comparison to more established pronunciation quality methods (DTW-based path costs and forced alignment of canonical transcription) using the same data sets is an on-going work. Future work includes also solving the code-switching issue for the CTC training.



\bibliographystyle{IEEEtran}

\bibliography{ref}


\end{document}